\documentclass[
]{ceurart}

\usepackage{listings}
\usepackage{colortbl}
\usepackage{amsmath}
\begin{document}

\copyrightyear{2023}
\copyrightclause{Copyright for this paper by its authors. Use permitted under Creative Commons License Attribution 4.0 International (CC BY 4.0)}

\conference{Forum for Information Retrieval Evaluation, December 15-18, 2023, India}

\title{Software Metadata Classification based on Generative Artificial Intelligence}


\author[1]{Seetharam Killivalavan}[%
orcid=0009-0002-2655-117X,
email=seetharam2210463@ssn.edu.in
]
\cormark[1]
\fnmark[1]
\address[1]{Sri Sivasubramaniya Nadar College of Engineering,
  Chennai, Tamil Nadu- 603110}
  
\author[2]{Durairaj Thenmozhi}[%
orcid=0000-0003-0681-6628,
email=theni_d@ssn.edu.in
]
\cormark[1]
\fnmark[1]

\cortext[1]{Corresponding author.}

\begin{abstract}
  This paper presents a novel approach to enhance the performance of binary code comment quality classification models through the application of Generative Artificial Intelligence (AI). By leveraging the OpenAI API, a dataset comprising 1239 newly generated code-comment pairs, extracted from various GitHub repositories and open-source projects, has been labelled as \textit{"Useful"} or \textit{"Not Useful"}, and integrated into the existing corpus of 9048 pairs in the C programming language.
Employing a cutting-edge Large Language Model Architecture, the generated dataset demonstrates notable improvements in model accuracy. Specifically, when incorporated into the Support Vector Machine (SVM) model, a 6\% increase in precision is observed, rising from 0.79 to 0.85. Additionally, the Artificial Neural Network (ANN) model exhibits a 1.5\% increase in recall, climbing from 0.731 to 0.746.
This paper sheds light on the potential of Generative AI in augmenting code comment quality classification models. The results affirm the effectiveness of this methodology, indicating its applicability in broader contexts within software development and quality assurance domains. The findings underscore the significance of integrating generative techniques to advance the accuracy and efficacy of machine learning models in practical software engineering scenarios.

\end{abstract}

\begin{keywords}
  Code Comment Quality Classification \sep
  Generative Artificial Intelligence \sep
  Support Vector Machines \sep
  Artificial Neural Networks \sep
  Natural Language Processing 
\end{keywords}

\maketitle

\section{Introduction}
Code comments play a vital role in software development, aiding comprehension, collaboration and maintenance \cite{de2005study}. Manual evaluation is arduous, time-consuming and subjective \cite{haouari2011good}. We propose the integration of Generative AI to enhance automation \cite{ebert2023generative}, potentially revolutionizing code quality assessment and streamlining the\textit{ Software Development Life Cycle} (SDLC).
{
Facilitating an efficient SDLC, where comments guide developers, accelerating issue resolution and laying a strong foundation for future iterations \cite{majumdar2020comment}. This paper outlines our methodology, experiments and transformative potential for the software engineering community \cite{roehm2012professional}. The subsequent sections will provide a brief overview of the existing landscape of code comment classification and the generation of our dataset utilizing Large Language Models (LLMs).

\subsection{	CODE COMMENT CLASSIFICATION: CURRENT LANDSCAPE AND CHALLENGES}

In software development, code comments play a vital role in providing insights into logic, design choices, and potential challenges \cite{rani2021identify}. However, manual evaluation is inherently subjective, tedious and often inconsistent \cite{majumdar2020comment}. To address this, code comment classification has emerged, automating the categorization process, which entails assigning labels such as "Useful" or "Not Useful" to comments, streamlining codebase analysis \cite{majumdar2022automated}. In this paper, we explore the enhancement of classification models with Generative AI \cite{ebert2023generative}, potentially revolutionizing comment utility assessment. Automated classification allows for more efficient resource allocation, ensuring crucial comments receive due attention. This introduction sets the stage for a deeper examination of the impact of LLMs in shaping the future of code comment classification, demonstrating potential transformative progress in software development practices \cite{de2005study}.
\subsection{IMPACT OF LLM ON THE QUALITY OF COMMENTS}

The integration of Large Language Models (LLMs) represents a pivotal advancement in the
realm of code comment quality assessment \cite{ebert2023generative}. These models go beyond syntax, understanding
code’s semantic context. They generate relevant comments, streamlining the comment creation
process, which enhances utility for developers throughout the SDLC. The impact of LLMs extends
beyond classification, fundamentally influencing how developers engage with code. Integrating
LLMs promises more precise communication and more effective collaboration. This transformative
potential signifies the crucial role of LLMs in shaping the future of code comment quality
assessment.

\vspace{\baselineskip}
{
\raggedright
The integration of Generative AI in the IRSE@FIRE-2023 task \cite{majumdar2023generative}  poised to revolutionize code quality assessment and enhance the Software Development Life Cycle (SDLC), fostering more efficient resource allocation and collaboration among developers.

\vspace{\baselineskip}

The subsequent sections are structured as follows: Section \ref{sec2} delves into comment classification and Generative AI background. Section \ref{sec3} outlines the task and dataset. Section \ref{sec4} presents our methodology. Results are detailed in Section \ref{sec5}, and in Section \ref{sec6}, we compare our models and embeddings with existing approaches in code comment quality classification, highlighting their distinct contributions. Finally, Section \ref{sec7} concludes with a concise summary of our findings and a discussion of potential future research directions.
}

\section{Related Work \label{sec2}}
This section provides a comprehensive overview of prior research in code comment quality classification. Noteworthy studies have leveraged NLP and Text Analysis for classification \cite{rani2021identify}, addressed multilingual challenges with neural language models \cite{kostic2023monolingual} and introduced frameworks like CommentEval for accurate classification \cite{majumdar2022automated}. Other approaches introduced contextualized embeddings \cite{majumdar2022effective} and analyzed commenting habits in open-source Java projects \cite{haouari2011good}.

Additionally, Comment-Mine presented a semantic search architecture \cite{majumdar2020comment}, while ReposSkillMiner pioneered an approach leveraging the GitHub API and NLP \cite{kourtzanidis2020reposkillminer}. Prometheus introduced an efficient system for crawling and storing software repositories from GitHub \cite{jobst2022efficient}.

Our methodology stands out by incorporating Generative AI through Large Language Models (LLMs), offering a transformative approach to code comment quality classification. This novel methodology addresses existing limitations, promising a more accurate assessment of code comments.

Furthermore, our methodology, driven by generative AI and LLM, marks a significant stride in redefining code analysis and documentation in software development. This research paper showcases the transformative potential of advanced technologies in practical software engineering, presenting a new paradigm for code comment classification.

\section{Task and Dataset Description \label{sec3}}
This section outlines the IRSE@FIRE-2023 task \cite{majumdar2023generative}, focused on improving a binary code comment quality classification model. The task involves integrating newly generated code-comment pairs for enhanced accuracy. It comprises an initial dataset of 9048 labeled code-comment pairs in C, out of which 5378 were classified as "Useful" and 3670 were classified as "Not Useful", along with additional pairs generated using a Large Language Model (LLM), each labeled.

The desired output includes two versions of the classification model: one with the added generated pairs and labels, and another without. The starting dataset encompasses 9048 comments from GitHub, each with the comment text, surrounding code, and a corresponding usefulness label (Table \hyperlink{topoftab1}{1}).

\begin{table}[ht]
\hypertarget{topoftab1}{}
\caption{Sample Data Instance}
\centering
\label{tab1}
\begin{tabular}{|p{5cm}|p{5cm}|p{2.5cm}|}
\hline
{\cellcolor[rgb]{0.753,0.753,0.753}}\textbf{Comment}& {\cellcolor[rgb]{0.753,0.753,0.753}}\textbf{Code}& {\cellcolor[rgb]{0.753,0.753,0.753}}\textbf{Label}\\
\hline
/* Swap two values */& void swapValues(int *x, int *y) \{

 int temp; 

temp = *x; 
 *x = *y; 
  
  *y = temp;\}& Useful \\
\hline
/* Compute the sixth power */ & {int result = computeSixthPower(value);}& Not Useful \\
\hline
/*Simple variable declaration */& 
int x = 10;
& Not Useful \\
\hline
\end{tabular}

\end{table}

To establish the ground truth, 14 annotators assessed each comment independently, resulting in substantial agreement (Cohen’s kappa value of 0.734). The annotation process involved the assessment of a comprehensive set of  16,000 comments.

Participants are also tasked with generating an additional dataset of labeled code-comment pairs from GitHub using an LLM. This dataset is to be submitted alongside the task.

In summary, the objective is to refine the code comment quality classification model by integrating newly generated pairs, ultimately enhancing accuracy and effectiveness.

{
\raggedright
For further details, please refer to the task description provided at IRSE@FIRE-2023 \footnote{https://sites.google.com/view/irse2023/home}.
}

\section{Methodology \label{sec4}}
Our approach encompasses the combination robust methodologies, including Support Vector Machine (SVM) models for classification and Artificial Neural Networks (ANN) with diverse activation functions for capturing complex data relationships \cite{igual2017introduction}. Additionally, we leverage Large Language Models (LLMs) via the OpenAI API and utilize GitHub repositories to generate a diverse and substantial dataset of code-comment pairs. The following subtopics detail our specific methodologies: implementing SVM models, exploring ANN models, and generating datasets using the OpenAI API and GitHub repositories. These methodologies collectively form the foundation of our innovative approach to code comment quality assessment. Within the framework of our methodology, Figure \hyperlink{topofimg1}{1} elegantly elucidates the architectural blueprint that underpins our approach.

\begin{figure}[ht]
\hypertarget{topofimg1}{}
\centering
\includegraphics[width=\linewidth]{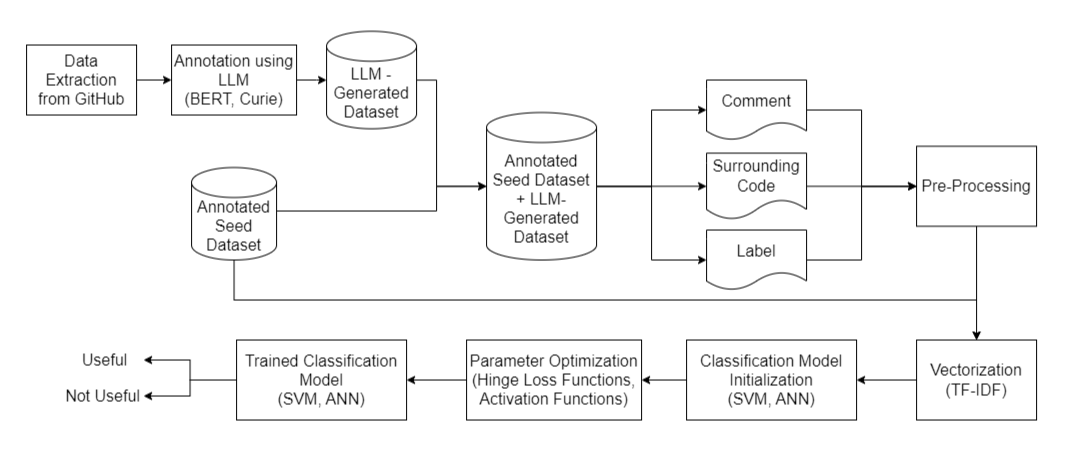}
\caption{Architecture Diagram}
\label{fig1}
\end{figure}

\subsection{Support Vector Machines}

A Linear Support Vector Machine (SVM) is a powerful classification technique that finds the optimal hyperplane for effective data separation, expressed as \(y = mx + b\), where \(y\) is the predicted class label, \(x\) is the input data, \(m\) is the slope and \(b\) is the y-intercept. It maximizes the margin, which is the distance between the hyperplane and the nearest data points. This margin (\textit{M}) can be calculated as:

\begin{equation}
M = \frac{2}{\|m\|} \tag{1}
\end{equation}

where ||\textit{m}|| is the length of the weight vector\textit{ m}.

{
\raggedright
SVM aims to minimize the square of the length of the weight vector (||\textit{m}||²) while ensuring that each data point $x_i$ is correctly classified:
}

\begin{equation}
y_i (m \cdot x_i + b) \geq 1 \tag{2}
\label{eqn2}
\end{equation}

{
\raggedright
Equation \ref{eqn2} states that the product must be greater than or equal to 1 for all data points, emphasizing the importance of well-defined class separation in SVM classification. This condition is central to SVM's goal of locating an optimal hyperplane, maximizing the margin, and guaranteeing accurate data point classification. Support vectors, those closest to the hyperplane, are pivotal in margin definition, thereby influencing SVM's overall performance.
}

\subsection{Artificial Neural Networks}
Artificial Neural Networks (ANNs) are adaptable machine learning models that draw inspiration from the architecture and operation of the human brain. They excel at discerning complex data relationships, making them highly effective for tasks like code comment quality classification.
The mathematical representation of a single neuron in an ANN is given by:

\begin{equation}
    Z = \textit{w}_1\textit{x}_1 + \textit{w}_2\textit{x}_2 + \ldots + \textit{w}_n\textit{x}_n + \textit{b} \tag{3}
\end{equation}

where $x_n$ are input features, $w_n$ are corresponding weights and $b$ is the bias term.

{
\raggedright
\vspace{\baselineskip}
The weighted sum (\textit{Z}) is then passed through an
activation function, which introduces non-linearity into the model. Different
activation functions yield different learning behaviours.

\vspace{\baselineskip}

Here are a few common activation functions and their
formulas:
\vspace{\baselineskip}
}

i) Logistic Function:
\[f(Z) = \frac{1}{1 + e^{-Z}} \tag{4}\]

ii) Rectified Linear Unit (ReLU):
\[f(Z) = \max(0, Z) \tag{5}\]

iii) Hyperbolic Tangent (tanh):
\[f(Z) = \frac{e^Z - e^{-Z}}{e^Z + e^{-Z}} \tag{6}\]
\subsection{Leveraging LLM for Generation of Dataset}
Our methodology encompasses a multi-step approach to dataset generation. Initially, we leveraged both the OpenAI API, powered by the Curie Model, and GitHub repositories to diversify our dataset. The API simulated real-world coding scenarios, producing authentic code-comment pairs and substantially augmenting our dataset. Complementing this, we extracted additional pairs from various open-source projects on GitHub, ensuring relevance and utility. This combined strategy significantly broadened the dataset's coverage while upholding high quality standards. Subsequently, the code-comment pairs underwent processing using OpenAI's Curie Model in conjunction with BERT for label generation, signifying comment usefulness. This involved presenting prompts with both code and comment, and employing the LLM to generate a label. Finally, the dataset was meticulously assembled, each entry comprising code, comment, and the corresponding generated label. This rigorous methodology serves as a robust foundation for our code comment quality classification model.

\section{Analysis of Results \label{sec5}}
The evaluation of our code comment quality classification model stands as a pivotal phase in gauging its efficacy. Leveraging a combination of Support Vector Machines (SVM) and Artificial Neural Networks (ANN) equipped with diverse activation functions (ReLU, identity, logistic, and tanh), we conducted a comprehensive analysis of the model's performance. This multifaceted approach not only provided us with a deeper understanding of the model's adaptability but also demonstrated its robustness across varying scenarios. Moreover, the integration of these methodologies yielded a marked improvement in the precision score. This enhancement underscores the model's heightened capacity to accurately categorize code comments based on their practical utility. These findings echo previous studies that have highlighted the efficacy of SVM and ANN models in code comment quality assessment. The utilization of various activation functions further emphasizes the versatility of our approach. Overall, this evaluation phase solidifies the effectiveness of our classification model and its potential applicability in real-world software development contexts.
\subsection{Classification Models}
The evaluation of our code comment quality classification models yielded insightful findings, showcasing the impact of integrating LLM-generated data into our seed dataset of 9048 entries. This initial dataset was thoughtfully partitioned into training, testing and validation sets, with the testing set comprising 1718 entries. 
With the Seed Data,  SVM exhibited commendable precision (0.79), while ANN with ReLU activation demonstrated remarkable effectiveness, resulting in a notable recall score (0.731). Models with tanh and logistic activation functions showed similar precision scores of 0.726 and 0.73.

Post integration of 1239 LLM-generated entries, which seamlessly enriched the Seed Data, SVM's precision notably increased by 6\%, elevating the preceding value to 0.85, highlighting the value of incorporating generative AI. Using ReLU, ANN achieved a noteworthy 1.5\% rise in its recall, giving it a final recall of 0.746, while tanh and logistic functions yielded marginal changes. Extensive experimentation with varied SVM models and ANN activation functions was performed, and the results depicts the effectiveness of our approach, emphasizing the importance of meticulous experimentation in fine-tuning models for code comment quality analysis.

Furthermore, for detailed numerical insights, please refer to Table \hyperlink{topoftab2}{2}, which provides a comparison of the model performance, offering the classification report of our top-performing models. It serves as a comprehensive reference for our findings and allows the comparison of test accuracies and F1 scores before and after integration.

\vspace{\baselineskip}

\begin{table}[ht]
\hypertarget{topoftab2}{}
\label{tab2}
\centering
\caption{Model Performance Comparison}
\begin{tabular}{|c|c|c|c|c|}
\hline
\rowcolor[rgb]{0.753,0.753,0.753}
\textbf{Model}& \multicolumn{2}{c|}{\textbf{Performance with Seed Data}} & \multicolumn{2}{c|}{\textbf{Performance with Integrated Data}} \\
\cline{2-5}
\rowcolor[rgb]{0.753,0.753,0.753} & \textbf{Test Accuracy} & \textbf{F1-Score} & \textbf{Test Accuracy} & \textbf{F1-Score} \\
\hline
Linear SVM& 0.802 & 0.787& 0.811 & 0.789\\
\hline
SVM (poly. kernel) & 0.79 & 0.77 & 0.805 & 0.78 \\
\hline
ANN (ReLU) & 0.742 & 0.73& 0.75 & 0.74\\
\hline
ANN (tanh) & 0.74 & 0.725& 0.743& 0.733\\
\hline
ANN (logistic) & 0.739 & 0.721 & 0.741 & 0.73 \\
\hline
ANN (identity) & 0.741 & 0.724 & 0.742 & 0.729 \\
\hline
\end{tabular}
\end{table}

\subsection{Analysis of Dataset Generated using LLM}
The integration of data generated by OpenAI's Large Language Model (LLM), in conjunction with the utilization of the Curie model, and the inclusion of diverse datasets from various GitHub repositories and open-source projects represents a significant stride in elevating our code comment quality classification model. By meticulously adding 1239 new entries to our original dataset, we substantially enriched the diversity of our training corpus. This augmentation in data diversity led to a marked improvement in the accuracy of our classification model, benefiting both Support Vector Machine (SVM) and Artificial Neural Network (ANN) models. The heightened sensitivity achieved through this amalgamation enhances the model's generalization and prediction capabilities, underscoring the value of incorporating external data sources. Furthermore, the integration of BERT embeddings and the Curie model empowered our model to adeptly capture the intricacies of code commentary, notably enhancing its ability to distinguish between "Useful" and "Not Useful" comments. This capability proves crucial in real-world scenarios, where precise comment assessment plays a pivotal role in influencing the effectiveness of software development and maintenance processes.

\section{Discussion \label{sec6}}
In this section, we conduct a thorough comparative analysis of our models and embeddings in relation to previous studies on code comment classification. Our deliberate emphasis on Support Vector Machine (SVM) and Artificial Neural Network (ANN) models, each with specific activation functions, allows for an in-depth exploration of their efficacy. This focused investigation provides nuanced insights into their performance in code comment quality assessment, contrasting with the broader set of classifiers utilized by Majumdar et al. (2022a) \cite{majumdar2022automated}.

Additionally, our research methodology diverges from the work of Majumdar et al. (2020) \cite{majumdar2020comment}, which primarily centers on the extraction of knowledge domains from code comments for addressing developer queries during maintenance. In contrast, our focus centers on the development and evaluation of code comment quality classification models. This includes the integration of LLM-generated data, resulting in significant enhancements in classification precision.

Concerning embeddings, Majumdar et al. (2022b) \cite{majumdar2022effective} emphasize contextualized word representations fine-tuned on software development texts. In our case, we utilized both BERT and custom embeddings specifically tailored for software development concepts. This approach provided high-dimensional semantic representations, catering to a wide array of natural language processing tasks. It's worth noting that for labeling, we harnessed the Curie model. This distinction underscores the versatility and broader applicability of our embeddings compared to the contextualized embeddings discussed by Majumdar et al (2022b)\cite{majumdar2022effective}.

Fundamentally, our proposition emphatically focuses on specific models and embeddings, providing unique insights into their effectiveness for assessing code comment quality. The emphasis on specific models and customized embeddings offers detailed insights into evaluating code comment quality, distinguishing it from the broader, contextually-focused techniques utilized in prior research \cite{majumdar2022effective}.

\section{Conclusion \label{sec7}}
In conclusion, our research marks a significant advancement in code comment classification, particularly in binary classification. This progress was achieved through the strategic use of state-of-the-art machine learning models. We leveraged the strengths of Support Vector Machine (SVM) and Artificial Neural Network (ANN) models, each excelling in their respective classification tasks. The SVM model, known for its ability to define clear decision boundaries, played a pivotal role. Meanwhile, the ANN model, inspired by the human brain's workings, proved invaluable in capturing intricate data relationships. Our exploration of various activation functions for the ANN further underscores the thoroughness of our approach. This combination of methodologies greatly contributed to our classification success, setting the stage for more nuanced approaches in software development and quality assurance.

\vspace{\baselineskip}
{

The integration of data generated through OpenAI's Large Language Models proved instrumental, resulting in the creation of 1239 new data entries. This augmentation resulted in a commendable 6\% increase in precision and a notable 1.5\% boost in recall for the SVM and ANN models respectively. Prior to the integration of LLM-generated data, our SVM and ANN models demonstrated test accuracies of 80.2\% and 74.2\% respectively. Post-integration, these figures rose to 81.1\% and 75\%, underscoring the tangible impact of this augmentation.

\vspace{\baselineskip}

In the near future, the implications of this research paper extend beyond the confines of code comment classification. The methodologies employed here have laid a foundation for more nuanced and adaptable approaches to tasks in software development and quality assurance. The integration of generative AI, as demonstrated through the use of LLM, holds promise in revolutionizing how we approach code analysis and documentation. As software landscapes continue to evolve, this paper serves as a testament to the transformative potential of incorporating cutting-edge technologies in practical software engineering scenarios.
}

\bibliography{sample-ceur}


\end{document}